\RequirePackage{lineno} 

\documentclass[twocolumn,preprintnumbers,amsmath,amssymb,nofootinbib]{revtex4}

\usepackage{graphicx}
\usepackage{dcolumn}
\usepackage{bm}
 
\usepackage{epstopdf}

\def\bea{\begin{eqnarray}}
\def\eea{\end{eqnarray}}

\def\pp{\mbox{$p$-$p$}}

\def\ppb{\mbox{$p$-Pb}}

\def\pn{\mbox{$p$-N}}

\def\nn{\mbox{N-N}}

\def\pt{$p_t$}

\def\yt{$y_t$}

\def\nch{$n_{ch}$}
\def\mmpt{$\bar p_t$}

\begin{document} 

\setlength{\pdfpagewidth}{8.5in}
\setlength{\pdfpageheight}{11in}

\setpagewiselinenumbers
\modulolinenumbers[5]

\addtolength{\footnotesep}{-10mm}\

\preprint{version 0.4\textsl{}}

\title{Mass-dependent transport of hadron species from soft to hard (nonjet to jet) spectrum components within small collision systems at the large hadron collider
}

\author{Thomas A.\ Trainor}\affiliation{University of Washington, Seattle, WA 98195}


\date{\today}

\begin{abstract}
In previous analyses a two-component (soft+hard) model (TCM) was developed for identified-hadron (PID) spectra from 5 TeV $p$-Pb and 13 TeV $p$-$p$ collisions. Spectrum data are generally described within their statistical uncertainties. Within the model are coefficients $z_{si}(n_s)$ and $z_{hi}(n_s)$ that denote the fractions of hadron species $i$ within total soft $\bar \rho_s$ and hard $\bar \rho_h$ charge densities and that vary  significantly with event index $n_s = \Delta \eta \bar \rho_s$. This letter reports that variation of those coefficients with $n_s$ implies transport of hadron species from soft component to hard component, increasingly with increased jet production, while conserving the total particle number for each species that is predicted by a statistical model. The extent of transport is simply proportional to hadron mass. 
\end{abstract}


\maketitle

Since commencement of high-energy nuclear collisions at the large hadron collider (LHC) certain data features have been associated with ``collectivity'' (flows) in small collision systems -- \pp\ and \ppb\ -- including one or two ``ridges'' in two-particle angular correlations~\cite{ppcms,ppbridge} and certain trends for identified-hadron (PID) \pt\ spectra~\cite{aliceppbpid,alicepppid}. Such data interpretations have lead to dramatic claims -- recently of an ongoing ``...revolution...driven by the experimental observation of flow-like features in the collisions of small hadronic systems''~\cite{nagle}. Such claims have motivated detailed examination of PID \pt\ spectra from 5 TeV \ppb\ spectra~\cite{ppbpid,pidpart1,pidpart2} and 13 TeV \pp\ spectra~\cite{pppid} via a two-component (soft+hard) model (TCM). The TCM describes PID spectra for both collision systems within their statistical uncertainties and no evidence for radial flow is apparent. Hadron production is dominated by two processes: projectile nucleon dissociation (soft) and large-angle parton scattering to jets (hard). What has also emerged from TCM analysis is an apparent mass-dependent transport of hadron species from nonjet to jet component which presents interesting implications for collision mechanisms in small collision systems. That is the main subject of this letter.

A TCM description of PID spectrum data relies on previous establishment of a TCM for nonPID spectra from the same collision system. For \pp\ collisions the TCM was established and developed in Refs.~\cite{ppprd,alicetommpt,alicetomspec,tomnewppspec}. For \ppb\ collisions a TCM  was established in Refs.~\cite{tommpt,tomglauber}. A central issue for \ppb\ TCM definition is the collision geometry or centrality by which trends for $\bar \rho_s$ and $\bar \rho_h$ are determined from measured charge density $\bar \rho_0$. Determination of 5 TeV \ppb\ centrality is reported in Ref.~\cite{tomglauber}.

A TCM for identified hadrons may be generated by assuming that each hadron species $i$ comprises certain {\em fractions} of TCM soft- and hard-component charge densities $\bar \rho_s$ and $\bar \rho_h$ (summing to total charge density $\bar \rho_0$) denoted by $z_{si}$ and $z_{hi}$. The PID spectrum TCM is then
\bea \label{pidspectcm}
\bar \rho_{0i}(y_t,n_s)
&\approx&  z_{si}(n_s) \bar \rho_{s} \hat S_{0i}(y_t) +   z_{hi}(n_s) \bar \rho_{h} \hat H_{0i}(y_t).~~
\eea
For \pp\ spectra $\bar \rho_h \approx \alpha \bar \rho_s^2$, where $\alpha(\sqrt{s})$ is $O(0.01)$ and slowly varying with collision energy~\cite{alicetomspec}. For \mbox{A-B} collisions, with $N_{part}/2$ nucleon (N) participant pairs and $N_{bin}$ \mbox{N-N} binary collisions, $\bar \rho_s = (N_{part}/2) \bar \rho_{sNN}$ and $\bar \rho_h = N_{bin} \bar \rho_{hNN}$. The hard/soft ratio is $\bar \rho_h / \bar \rho_s = x(n_s) \nu(n_s)$, where $x(n_s) = \bar \rho_{hNN}/\bar \rho_{sNN} \approx \alpha \bar \rho_{sNN}$ and $\nu(n_s) \equiv 2 N_{bin} / N_{part}$. For \pp\ collisions $\nu \rightarrow 1$.
Unit-normal model functions $\hat S_{0i}(y_t)$ and $\hat H_{0i}(y_t)$ are determined for each hadron species, but close correspondence with unidentified-hadron models is observed~\cite{ppbpid,pidpart1,pidpart2,pppid}. 

To estimate coefficients $z_{si}(n_s)$ spectra are rescaled as
\bea \label{xi}
 X_i(y_t,n_s) &\equiv& \frac{\bar \rho_{0i}(y_t,n_s)}{ \bar \rho_{si}(n_s)}
\nonumber \\
&\approx&  \hat S_{0i}(y_t) +  \tilde z_i(n_s) x(n_s)\nu(n_s) \hat H_{0i}(y_t),
\eea
where $\tilde z_i(n_s) \equiv z_{hi}(n_s)/z_{si}(n_s)$ and the rescale factor is $ \bar \rho_{si}(n_s) = z_{si}(n_s)\bar \rho_{s}$. In Ref.~\cite{ppbpid}, reporting TCM analysis of PID spectra from 5 TeV \ppb\ collisions, the expression
\bea \label{zsieq}
z_{si}(n_s) &=& z_{0i} \frac{1 - x(n_s)\nu(n_s)}{1 - \tilde z_i(n_s) x(n_s)\nu(n_s)}
\eea
with $z_{0i}$ defined by $\bar \rho_{0i}(n_s) = z_{0i} \bar \rho_{0}(n_s)$ (charge densities) was used with Eq.~\ref{xi} to determine $z_{0i}$ and $\tilde z_i$ as {\em fixed} values satisfying the condition that at low \yt\ $ X_i(y_t,n_s) \rightarrow  \hat S_{0i}(y_t)$. Given those parameter values quantity $z_{hi}(n_s) = \tilde z_i z_{si}(n_s)$ was predicted, leading to observation of substantial pion excess and proton inefficiency. In Ref.~\cite{pidpart1} the proton inefficiency was corrected, quantities $z_{si}(n_s)$ and $z_{hi}(n_s)$ were inferred separately from PID spectrum data, and ratio $\tilde z_i(n_s) = z_{hi}(n_s) / z_{si}(n_s)$ was then observed to have a significant event-class (i.e.\ $n_s = \Delta \eta \bar \rho_s$) dependence. However, the pion excess was overlooked.

In Ref.~\cite{pppid} the analysis strategy from Ref.~\cite{pidpart1} was applied to 13 TeV \pp\ collisions, with equivalent correction of proton spectra. In that case the pion excess issue was recognized, and a strategy to define pion $z_{si}(n_s)$ and $z_{hi}(n_s)$ on the basis of charge conservation among charged hadrons was adopted. However, spectrum data were of insufficient quality to infer variation of $\tilde z_i(n_s)$.

\begin{figure}[h]
	\includegraphics[width=3.3in]{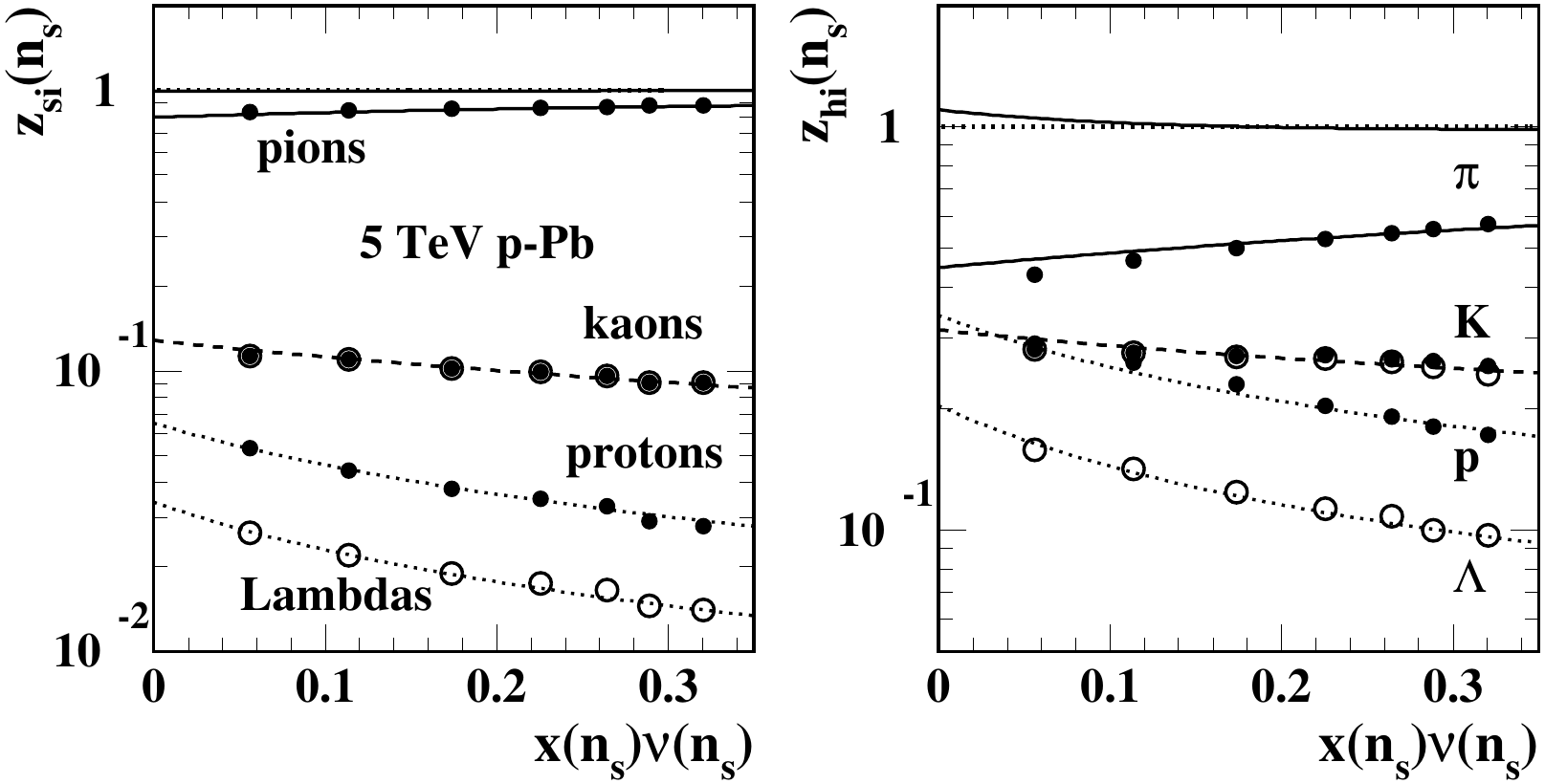}
	\put(-138,90) {\bf (a)}
	\put(-17,105) {\bf (b)}\\
	\includegraphics[width=3.3in]{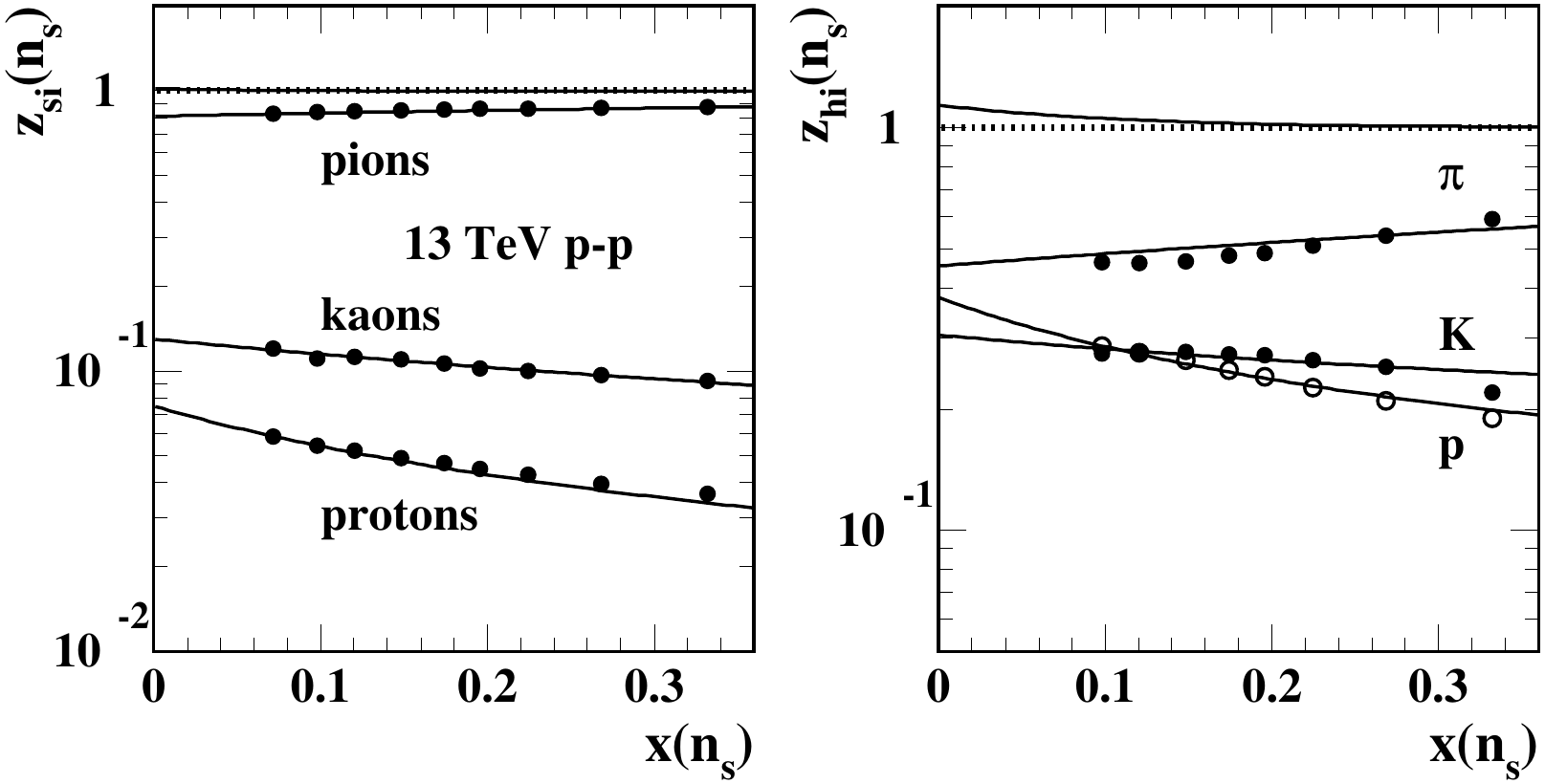}
	\put(-138,90) {\bf (c)}
	\put(-17,105) {\bf (d)}
	\caption{\label{zxs}
Trends vs hard/soft ratio $x(n_s)\nu(n_s)$ for soft and hard fractions $z_{si}(n_s)$ (left) and $z_{hi}(n_s)$ (right) and for 5 TeV \ppb\ collisions (a,b) and 13 TeV \pp\ collisions (c,d). In (a,b) charged hadrons (solid dots) are differentiated from neutral hadrons (open circles). The curves are derived from Eq.~(\ref{zsieq}) with parameters as described in the text. The uppermost solid curves are sums of lower curves for charged hadrons. In (a) the $z_{si}(n_s)$ values for $K^\pm$ are copied from those for $K_\text{S}^0$ ($\times 2$) due to the limited \pt\ acceptance of the former. 
	} 
\end{figure}

Figure~\ref{zxs} summarizes the values (points) for $z_{si}(n_s)$ (left) and $z_{hi}(n_s)$ (right) from 5 TeV \ppb\ collisions (a,b) and 13 TeV \pp\ collisions (c,d) plotted against hard/soft ratio $x(n_s)\nu(n_s)$ (with $\nu(n_s) \rightarrow 1$ for \pp\ collisions). It is important to note that measured $z_{si}(n_s)$ values are derived from PID spectra based only on the low-\yt\ structure of $\hat S_{0i}(y_t;T,n)$ and independently determined $\bar \rho_s$. Equation~(\ref{zsieq}) parametrizing $z_{si}(n_s)$ is a hypothesis based on assumptions (i.e.\ $z_{0i}$ and $\tilde z_{i}(n_s)$ are assumed approximately constant) that happens to describe inferred $z_{si}(n_s)$ data well. The pion values for $z_{si}(n_s)$ and $z_{hi}(n_s)$ from both 5 TeV \ppb\ collisions and 13 TeV \pp\ collisions are based on charge conservation relative to the other charged hadron species. The \ppb\ values for pion $z_{hi}(n_s)$ are thereby substantially reduced from those reported in Ref.~\cite{pidpart1}. The curves correspond to Eq.~(\ref{zsieq}) with $z_{0i}$ and $\tilde z_i(n_s)$ as defined in Ref.~\cite{pidpart1} except that pion values for parameters (A,B) in its Fig.~8 (right) are altered as (0.78,0.35) $\rightarrow$ (0.56,0.25) to accommodate the revised pion $z_{hi}(n_s)$ values satisfying charge conservation. As a consequence, hadron fractions from two collision systems and two energies are described by the same parametrization wherein $z_{0i}$ values are predicted by a statistical model~\cite{statmodel} and $\tilde z_i(n_s)$ values are simply proportional to hadron mass~\cite{pidpart2}.

While the trends in Fig.~\ref{zxs} combined with Eq.~(\ref{pidspectcm}) accurately describe PID spectra for two small collision systems at two energies an alternative presentation format suggests deeper implications. The alternatives are ratios of soft and hard charge densities to total charge density for each hadron species $i$ expressed as
\bea \label{simple}
\frac{\bar \rho_{si}(n_s)}{\bar \rho_{0i}(n_s)} &=& \frac{n_{si}}{n_{chi}} = \frac{1}{z_{0i}} \frac{z_{si}(n_s)}{(1 + x\nu)} = \frac{1}{1 + \tilde z_i x\nu}
\\ \nonumber
\frac{\bar \rho_{hi}(n_s)}{\bar \rho_{0i}(n_s)} &=& \frac{1}{z_{0i}} \frac{z_{hi}(n_s)x\nu}{(1 + x\nu)} = \frac{\tilde z_i x\nu}{1 + \tilde z_i x\nu},
\eea
where the rearranged ratios have quite simple forms. Those ratios eliminate statistical-model fractions $z_{0i}$ from the data presentation as a matter of simplification.

\begin{figure}[h]
	\includegraphics[width=3.3in]{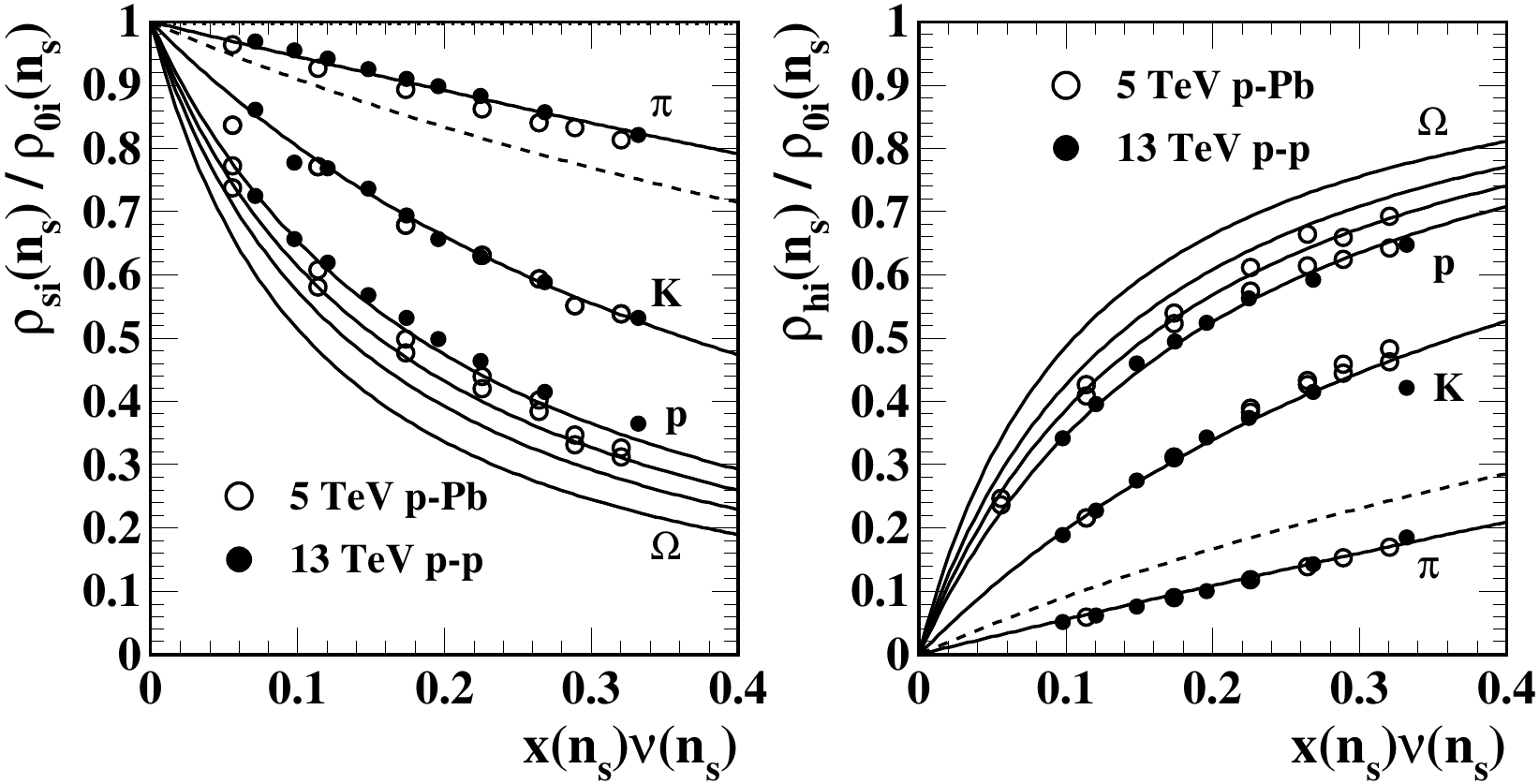}
	\caption{\label{5c}
Data in the form of Eqs.~(\ref{simple}) derived from $z_{si}(n_s)$ (left) and $z_{hi}(n_s)$ (right) and for 13 TeV \pp\ collisions (solid dots) and 5 TeV \ppb\ collisions (open circles). The solid curves are defined by Eqs.~(\ref{simple}) (rightmost expressions) while the dashed curves represents nonPID trends with $\tilde z_i = 1$. The six solid curves represent $\pi^\pm$, $K^\pm$ {\em and} $K_\text{S}^0$, $p$, $\Lambda$, $\Xi$ and $\Omega$.
} 
\end{figure}

Figure~\ref{5c} shows results from Eqs.~(\ref{simple}) for several hadron species from two collision systems based on $z_{si}(n_s)$ and $z_{hi}(n_s)$ data from Fig.~\ref{zxs}. The solid curves depend only on the universal $\tilde z_i(n_s)$ trends reported in Ref.~\cite{pidpart1}. The dashed curves correspond to nonPID $x\nu$ trends with $\tilde z_i = 1$. Data for 5 TeV \ppb\ collisions are open circles; data for 13 TeV \pp\ are solid dots. The \ppb\ data include neutral kaons $K^0_\text{S}$ and Lambdas $\Lambda$. Trends for Cascades $\Xi$ and Omegas $\Omega$ are predicted based on extrapolation of $\tilde z_i$ hadron mass dependence reported in Ref.~\cite{pidpart1}. For high-multiplicity \pp\ and \ppb\ collisions the predicted  $\Omega$ hard-component (jet-fragment) fraction is near 80\%. This data-model comparison demonstrates that hadrochemistry in small collision systems at the LHC is determined by two fixed parameters defining $\tilde z_i(n_s)$ and statistical-model predictions of $z_{0i}$ that predate LHC startup.

To summarize those results, as nonPID hard/soft (jet/nonjet) ratio $x\nu$ increases, the fractions of more-massive hadrons (compared to pions) in the hard component increase while those increases are exactly compensated by decreased fractions in the soft component. In effect, with increasing jet production hadrons of species $i$ are {\em transported} from the spectrum component associated with projectile-nucleon dissociation to the spectrum component associated with jet fragments. The transport conserves the total number of hadrons of species $i$ represented by fractions $z_{0i}$ predicted by a statistical model.

Given the PID spectrum TCM in Eq.~(\ref{pidspectcm}) the corresponding ensemble-mean {\em total} \pt\ for identified hadrons of species $i$, integrated over some angular acceptance $\Delta \eta$ that includes integrated charge $n_{chi}$, is
\bea \label{ptintid}
\bar P_{ti} 
&=&  z_{si}(n_s)  n_{s} \bar p_{tsi} +  z_{hi}(n_s) n_{h} \bar p_{thi}(n_s).
\eea 
The TCM for conventional ensemble-mean $\bar p_{ti}$  is then
\bea \label{pampttcmid}
\frac{\bar P_{ti}}{n_{chi}}  &\equiv&    \bar p_{ti}   \approx \frac{\bar p_{tsi} + \tilde z_i(n_s) x(n_s)\nu(n_s) \, \bar p_{thi}(n_s)}{1 +  \tilde z_i(n_s)x(n_s)\nu(n_s)}.
\eea
As an alternative, a \mmpt\ expression based on the TCM -- see Eq.~(\ref{simple}) (upper) --  has the particularly simple form
\bea \label{pampttcmpid}
\frac{\bar P_{ti}}{n_{si}} &=& \frac{n_{chi}}{n_{si}}\bar p_{ti} \approx \bar p_{tsi} + \tilde z_i(n_s) x(n_s)\nu(n_s) \, \bar p_{thi}(n_s).
\eea
Factors $\tilde z_i(n_s) x(n_s)\nu(n_s)$ have been defined above and soft- and hard-component mean values $\bar p_{tsi}$ and $\bar p_{thi}(n_s)$ are obtained from TCM models $\hat S_{0i}(y_t)$ and $\hat H_{0i}(y_t,n_s)$, where $\hat H_{0i}(y_t,n_s)$ may be slowly varying with $n_s$~\cite{pidpart2,pppid}.

\begin{figure}[h]
		\includegraphics[width=1.65in]{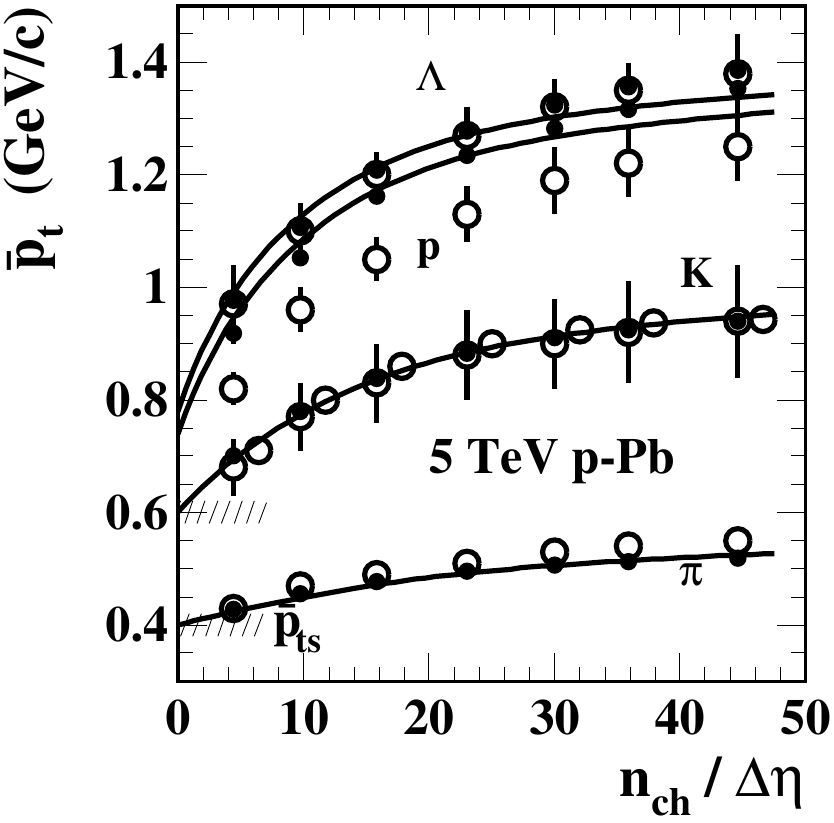}
		\includegraphics[width=1.65in]{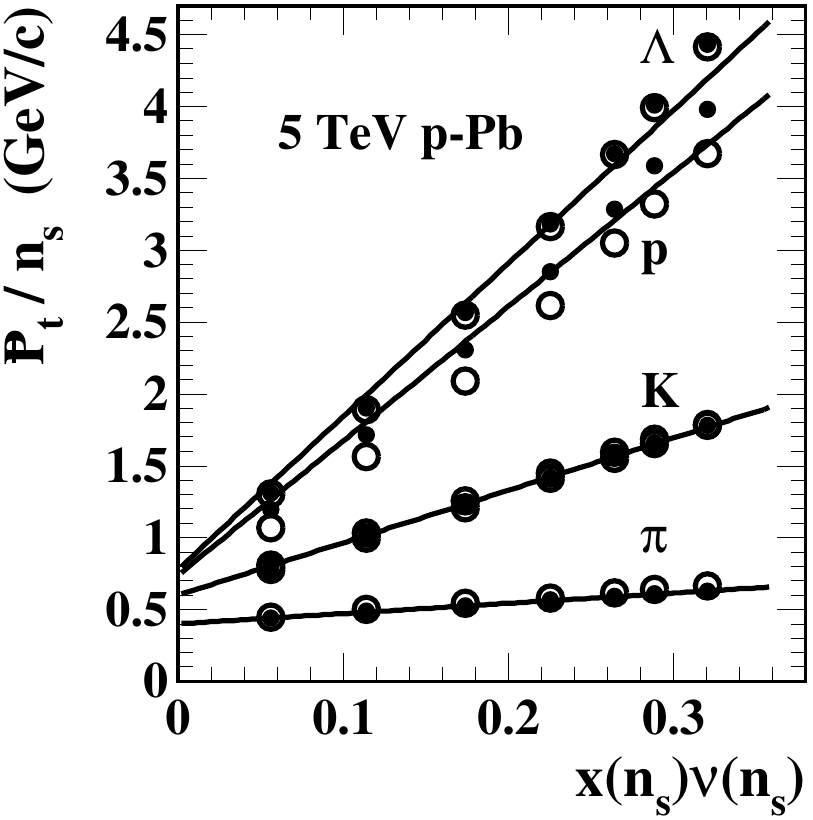}
	\put(-143,85) {\bf (a)}
\put(-19,80) {\bf (b)}\\
	\includegraphics[width=3.3in]{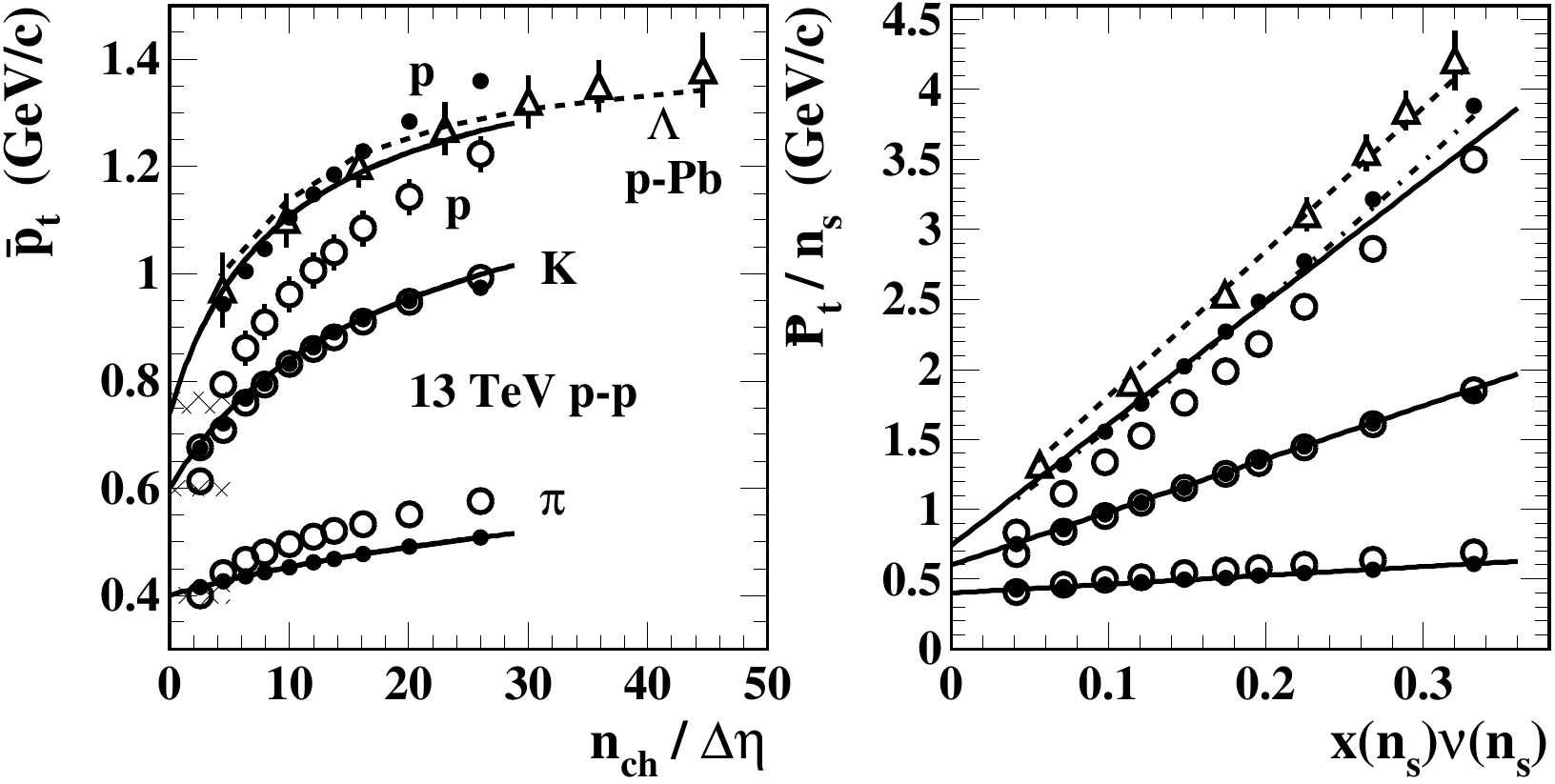}
	\put(-141,80) {\bf (c)}
\put(-20,80) {\bf (d)}
	\caption{\label{mpttrends}
Left: \mmpt\ data (open circles) for 5 TeV \ppb\ collisions (a) from Ref.~\cite{aliceppbpid} and for 13 TeV \pp\ collisions (c) from Ref.~\cite{alicepppid}. Solid dots are full TCM evaluated with \nch\ on data points. The curves are Eq.~(\ref{pampttcmid}) with fixed $\bar p_{thi}(n_s)$ for event class 4.		
 Right: Results in the left panels replotted in the format of Eq.~(\ref{pampttcmpid}) revealing approximately linear trends on $x(n_s)\nu(n_s)$. In this TCM format the equivalence of two collision systems differing in size and collision energy is apparent.
	} 
\end{figure}

Figure~\ref{mpttrends} shows \mmpt\ data (open circles) for 
5 TeV \ppb\ collisions from Ref.~\cite{aliceppbpid} (a) and for 13 TeV \pp\ collisions from Ref.~\cite{alicepppid} (c) plotted in the format of Fig.~7 of that reference. The solid dots are derived from TCM spectrum models (including for corrected proton spectra). The open triangles in (c) are Lambda data for 5 TeV \ppb\ collisions as reported in Ref.~\cite{aliceppbpid} for comparison. The solid curves are Eq.~(\ref{pampttcmid}) with fixed $\bar p_{tsi}$ and $\bar p_{thi}(n_s)$ for event-class 4. The pion value $\bar p_{tsi}\approx 0.40$ GeV/c includes a resonance contribution. For \pp\ data in (c) $ x\nu \rightarrow x$ in Eq.~(\ref{pampttcmid}). Published pion \mmpt\ data (lowest open points in (a, c) fall substantially {\em above} the TCM predictions (lowest solid curves), which may relate to proton detection inefficiency {\em or pion-proton misidentification}~\cite{pppid}.

Figure~\ref{mpttrends} (b,d) shows the same data and curves in the format of Eq.~(\ref{pampttcmpid}) with trends vs hard/soft (jet/nonjet) ratio $x(n_s)\nu(n_s)$ or $x(n_s)$ exhibiting approximately linear dependences. Note that in this format Lambda data in (b,d) correspond closely to {\em corrected} proton data (solid dots). Also note in (b,d) that baryon (including corrected proton) \mmpt\ increases somewhat faster than the TCM linear reference (solid lines) due to hard-component shifts to higher \yt\ with increasing \nch\ as reported in Refs.~\cite{pidpart2,pppid}. The dash-dotted curve in (d) is the proton TCM with {\em variable} $\hat H_{0i}(y_t,n_s)$. The simplicity of this format demonstrates the importance of a TCM reference for precision tests of data trends and model quality. However, integral measures such as total $\bar P_{ti}$ and its ratio to $n_{chi}$ or $n_s$ represent {\em reduction} of information compared to intact differential PID spectra and their TCM representations.

Reference~\cite{alicepppid} offers several observations on light-flavor hadrochemistry concerning collision-energy dependence and multiplicity (\nch) dependence of collision systems. Comparing PID hadron production in \pp\ and \ppb\ collisions (at different collision energies) it is stated that ``...particle production scales with $dN_{ch}/d\eta$ [$\bar \rho_0$] independent of the colliding system.'' It is then asserted that, given recent 13 TeV \pp\ spectrum data, ``...for the first time, in pp collisions, we can disentangle the effect of center-of-mass energy from the multiplicity dependence of [three hadron species] production in a wide $p_T$ range.''  ``The $p_T$-integrated hadron-to-pion ratios as a function of multiplicity show no center-of-mass [energy] dependence and [those for \pp\ collisions at 13 TeV] are  compatible with [several other collision systems].'' It is concluded that ``...at the LHC energies, the chemical composition of primary hadrons scales with charged-particle multiplicity density  [$\bar \rho_0$] in a uniform way, despite the colliding system and collision energy.'' 

Those qualitative assertions may be compared with the energy-independent hadron abundance systematics vs hard/soft (jet/nonjet) ratio $x(n_s)\nu(n_s)$ -- {\em not} charge density $\bar \rho_0$ -- appearing in Fig.~\ref{5c} that represent those data accurately, and well-understood energy-dependence of model functions $\hat S_0(y_t)$ and $\hat H_0(y_t)$ that together describe LHC PID spectrum data within their statistical uncertainties~\cite{ppbpid,pidpart1,pidpart2,pppid}. TCM analysis of \mmpt\ data in Fig.~\ref{mpttrends} then emphasizes the great contrast between a conventional format in panels (a,c) and the TCM format in panels (b,d) where subtle data trends are accessible relative to a linear reference system. Note that in panel (c) based on ``scaling'' with charge density $\bar \rho_0$, Lambda data from 5 TeV \ppb\ collisions look very different from (corrected) proton data from 13 TeV \pp\ collisions whereas in panel (d) based on hard/soft ratio $x(n_s)\nu(n_s)$ the trends are quantitatively comparable, revealing mass-dependent variation of hard/soft abundance parameter $\tilde z_i(n_s)$.


TCM analysis distinguishes energy dependence of soft and hard model functions $\hat S_{0i}(y_t)$ and $\hat H_{0i}(y_t,n_s)$ for several hadron species relevant to differential spectra and energy dependence of parameter $\alpha(\sqrt{s_{NN}})$ in centrally-important relation $\bar \rho_{hNN} \approx \alpha(\sqrt{s_{NN}}) \bar \rho_{sNN}^2$ from the energy {\em independence} of PID fraction coefficients $z_{si}(n_s)$ and $z_{hi}(n_s)$ that control integrated PID yields. $z_{si}(n_s)$ applies to $p_t < 0.5$ GeV/c where there is no significant jet contribution, whereas $z_{hi}(n_s)$ determines the spectrum amplitude for $p_t \sim 10$ GeV/c and above that is certainly entirely jet fragments. The figures presented here demonstrate that PID spectrum and yield data are simply and {\em quantitatively} interpretable in the context of jet/nonjet (hard/soft) ratios  $\tilde z_i(n_s)$ and $x(n_s)\nu(n_s)$. Presentation of quantities in ratio to charge density $\bar \rho_0$ plotted vs  $\bar \rho_0$, such as Fig.~3 (a, c) confuse two distinct production mechanisms and are therefore practically uninterpretable.

Figure~\ref{5c} presents the main result of this letter. While the basic experimental facts are summarized just below that figure an attempt at further interpretation relies on details of \nn\ collision dynamics. Regarding  \pp\ inelastic collisions, at one extreme a nucleon projectile may dissociate along the beam axis (soft component), some elements (low-$x$ gluons) of the corresponding parton distribution function (PDF) may undergo large-angle scattering, and scattered gluons may then fragment to hadron jets consistent with measured fragmentation functions (hard component). One might then expect  PID {\em hadron} fractions of soft and hard components to be independent.

Figure~\ref{5c} instead implies that PID hadron production arising from projectile nucleon dissociation and from large-angle parton (gluon) scattering are {\em perfectly correlated} within data uncertainties. That result may be related to other characteristics of elementary collisions hinting at the importance of a quantum-mechanical description. Quantitative validity of the TCM for \pp\ collisions and its relation $\bar \rho_h \approx \alpha \bar \rho_s^2$ imply that each parton in one projectile may interact with {\em any} parton in the other projectile: geometric centrality is not relevant. Analysis of collision centrality in \ppb\ collisions~\cite{tomglauber} confirms that \pn\ collisions are ``all or nothing,'' with the concept of {\em exclusivity}: a projectile proton cannot interact with a different target nucleon {\em while engaged} in an inelastic \pn\ collision~\cite{tomexclude}. That conclusion is diametrically opposed to the classical Glauber model of A-B centrality~\cite{aliceglauber}.

One can then propose the following scenario for an inelastic \nn\ collision: An {\em event-wise} PDF develops {\em with hadronic properties} that reflect a statistical-model species distribution. The PDF depth on momentum fraction $x$ determines the event multiplicity, and especially the contribution from low-$x$ gluons relevant to midrapidity observations. Some fraction of low-$x$ gluons scatters to become jets (hard component), but the hadronic content of those jets is {\em perfectly complementary  to what remains} with the dissociated projectiles (soft component). Correlated hadron transport via jet formation leads to transport from higher $x$ ($y_z$) to higher $p_t$ producing the mass dependence of PID \mmpt\ trends. PID \ppb\ spectrum data suggest that each \pn\ collision proceeds as an isolated quantum transition (``all or nothing''); there is no ``cross talk'' among multiple \pn\ collisions, no collectivity. 

In summary, an identified-hadron (PID) two-component (soft+hard) model (TCM) implemented for two small collision systems -- 5 TeV \ppb\ collisions and 13 TeV \pp\ collisions -- describes PID spectrum data within their statistical uncertainties. Examination of TCM parameter trends reveals that variation of hadron species fractions $z_{si}(n_s)$ and $z_{hi}(n_s)$ are statistically equivalent for the two systems and further indicates mass-dependent transport from soft (nonjet) to hard (jet) component with increasing jet production that conserves  total hadron number for each species. Particle {\em sums} (soft+hard) for several hadron species follow statistical-model predictions that predate LHC startup.


\end{document}